\documentclass[runningheads,a4paper]{llncs}

\usepackage{amssymb,amsmath,bbm}
\usepackage{graphicx}

\usepackage{url}
\urldef{\mailjr}\path|juste.raimbault@polytechnique.edu|
\urldef{\mailjk}\path|julien.keutchayan@polytechnique.edu|
  
\newcommand{\keywords}[1]{\par\addvspace\baselineskip
\noindent\keywordname\enspace\ignorespaces#1}

\newcommand{\noun}[1]{\textsc{#1}}

\begin{document}

\mainmatter  

\title{A Discrepancy-based Framework to Compare Robustness between Multi-Attribute Evaluations}

\titlerunning{Discrepancy-based Framework}

%
%
\author{\noun{Juste Raimbault}$^{1,2}$} 
\authorrunning{Discrepancy-based Framework}

\institute{$^{1}$ UMR CNRS 8504 G{\'e}ographie-Cit{\'e}s, Paris, France\\
$^{2}$ UMR-T IFSTTAR 9403 LVMT, Champs-sur-Marne, France\\
\mailjr\\
}

\toctitle{Lecture Notes in Computer Science}
\tocauthor{Authors' Instructions}
\maketitle

\begin{abstract}
Multi-objective evaluation is a necessary aspect when managing complex systems, as the intrinsic complexity of a system is generally closely linked to the potential number of optimization objectives.
However, an evaluation makes no sense without its robustness being given (in the sense of its reliability). Statistical robustness computation methods are highly dependent of underlying statistical models. We propose a formulation of a model-independent framework in the case of integrated aggregated indicators (multi-attribute evaluation), that allows to define a relative measure of robustness taking into account data structure and indicator values. We implement and apply it to a synthetic case of urban systems based on Paris districts geography, and to real data for evaluation of income segregation for Greater Paris metropolitan area. First numerical results show the potentialities of this new method. Furthermore, its relative independence to system type and model may position it as an alternative to classical statistical robustness methods.
\end{abstract}

\section{Introduction}

\subsection{General Context}

Multi-objective problems are organically linked to the complexity of underlying systems. Indeed, either in the field of \emph{Complex Industrial Systems}, in the sense of engineered systems, where construction of Systems of Systems (SoS) by coupling and integration often leads to contradictory objectives~\cite{marler2004survey}, or in the field of \emph{Natural Complex Systems}, in the sense of non engineered physical, biological or social systems that exhibit emergence and self-organization properties, where objectives can e.g. be the result of heterogeneous interacting agents (see~\cite{newman2011complex} for a large survey of systems concerned by this approach), multi-objective optimization can be explicitly introduced to study or design the system but is often already implicitly ruling the internal mechanisms of the system. The case of socio-technical Complex Systems is particularly interesting as, following~\cite{haken2003face}, they can be seen as hybrid systems embedding social agents into ``technical artifacts'' (sometimes to an unexpected degree creating what \noun{Picon} describes as \emph{cyborgs}~\cite{picon2013smart}), and thus cumulate propensity to be at the origin of multi-objective issues\footnote{We design by \emph{Multi-Objective Evaluation} all practices including the computation of multiple indicators of a system (it can be multi-objective optimization for system design, multi-objective evaluation of an existing system, multi-attribute evaluation ; our particular framework corresponds to the last case).}. The new notion of \emph{eco-districts}~\cite{souami2012ecoquartiers} is a typical example where sustainability implies contradictory objectives. The example of transportation systems, which conception shifted during the second half of the 20th century from cost-benefit analysis to multi-criteria decision-making, is also typical of such systems~\cite{bavoux2005geographie}. Geographical system are now well studied from such a point of view in particular thanks to the integration of multi-objective frameworks within Geographical Information Systems~\cite{carver1991integrating}. As for the micro-case of eco-districts, meso and macro urban planning and design may be made sustainable through indicators evaluation~\cite{jegou2012evaluation}.

A crucial aspect of an evaluation is a certain notion of its reliability, that we call here \emph{robustness}. 
Statistics naturally include this notion since the construction and estimation of statistical models give diverse indicators of the consistence of results~\cite{launer2014robustness}. The first example that comes to mind is the application of the law of large numbers to obtain the \emph{p-value} of a model fit, that can be interpreted as a confidence measure of estimates. Besides, confidence intervals and \emph{beta-power} are other important indicators of statistical robustness. Bayesian inference provide also measures of robustness when distribution of parameters are sequentially estimated. Concerning multi-objective optimization, in particular through heuristic algorithms (for example genetic algorithms, or operational research solvers), the notion of robustness of a solution concerns more the stability of the solution on the phase space of the corresponding dynamical system. Recent progresses have been done towards unified formulation of robustness for a multi-objective optimization problem, such as~\cite{deb2006introducing} where robust Pareto-front as defined as solutions that are insensitive to small perturbations. In~\cite{1688537}, the notion of degree of robustness is introduced, formalized as a sort of continuity of other solutions in successive neighborhood of a solution.

However, there still lack generic methods to estimate robustness of an evaluation that would be model-independent, i.e. that would be extracted from data structure and indicators but that would not depend on the method used. Some advantages could be for example an \emph{a priori} estimation of potential robustness of an evaluation and thus to decide if the evaluation is worth doing. We propose here a framework answering this issue in the particular case of Multi-attribute evaluations, i.e. when the problem is made unidimensional by objectives aggregation. It is data-driven and not model-driven in the sense that robustness estimation does not depend on how indicators are computed, as soon as they respect some assumptions that will be detailed in the following.

\subsection{Proposed Approach}

\paragraph{Objectives as Spatial Integrals}

We assume that objectives can be expressed as spatial integrals, so it should apply to any territorial system and our application cases are urban systems. It is not that restrictive in terms of possible indicators if one uses suitable variables and integrated kernels : in a way analog to the method of geographically weighted regression~\cite{brunsdon1998geographically}, any spatial variable can be integrated against regular kernels of variable size and the result will be a spatial aggregation which sense depends on kernel size. The example we use in the following such as conditional means or sums suit well the assumption. Even an already spatially aggregated indicator can be interpreted as a spatial indicator by using a Dirac distribution on the centroid of the corresponding area.

\paragraph{Linearly Aggregated Objectives}

A second assumption we make is that the multi-objective evaluation is done through linear aggregation of objectives, i.e. that we are tackling a multi-attribute optimization problem. If $(q_i(\vec{x}))_i$ are values of objectives functions, then weights $(w_i)_i$ are defined in order to build the aggregated decision-making function $q(\vec{x})=\sum_i{w_i q_i(\vec{x})}$, which value determines then the performance of the solution. It is analog to aggregated utility techniques in economics and is used in many fields.
The subtlety lies in the choice of weights, i.e. the shape of the projection function, and various approaches have been developed to find weights depending on the nature of the problem. Recent work~\cite{dobbie2013robustness} proposed to compare robustness of different aggregation techniques through sensitivity analysis, performed by Monte-Carlo simulations on synthetic data. Distribution of biases where obtained for various techniques and some showed to perform significantly better than others. Robustness assessment still depended on models used in that work.

\bigskip

The rest of the paper is organized as follows : section 2 describes intuitively and mathematically the proposed framework ; section 3 then details implementation, data collection for case studies and numerical results for an artificial intra-urban case and a metropolitan real case ; section 4 finally discuss limitations and potentialities of the method.

\section{Framework Description}

\subsection{Intuitive Description}

We describe now the abstract framework allowing theoretically to compare robustnesses of evaluations of two different urban systems. Our framework is a generalization of an empirical method proposed in~\cite{ecodistrictReport} besides a more general benchmarking study on indicator sense and relevance in a sustainability context. Intuitively, it relies on empirical base resulting from the following axioms :
\begin{itemize}
\item Urban systems can be seen from the information available, i.e. raw data describing the system. As a data-driven approach, this raw data is the basis of our framework and robustness will be determined by its structure.
\item From data are computed indicators (objective functions). We assume that a choice of indicators is an intention to translate particular aspects of the system, i.e. to capture a realization of an ``urban fact'' (\emph{fait urbain}) in the sense of \noun{Mangin}~\cite{mangin1999projet} - a sort of stylized fact in terms of processes and mechanisms, having various realizations on spatially distinct systems, depending on each precise context.
\item Given many systems and associated indicators, a common space can be built to compare them. In that space, data represents more or less well real systems, depending e.g. on initial scale, precision of data, missing data. We precisely propose to capture that through the notion of point cloud discrepancy, which is a mathematical tool coming from sampling theory expressing how a dataset is distributed in the space it is embedded in~\cite{dick2010digital}. 
\end{itemize}

Synthesizing these requirements, we propose a notion of \emph{Robustness} of an evaluation that captures both, by combining data reliability with relative importance,
\begin{enumerate}
\item \emph{Missing Data} : an evaluation based on more refined datasets will naturally be more robust.
\item \emph{Indicator importance} : indicators with more relative influence will weight more on the total robustness.
\end{enumerate}

\subsection{Formal Description}

\paragraph{Indicators}

Let $(S_{i})_{1\leq i\leq N}$ be a finite number of geographically disjoints territorial systems, that we assume described through raw data and intermediate indicators, yielding $S_{i}=(\mathbf{X}_{i},\mathbf{Y}_{i})\in\mathcal{X}_{i}\times\mathcal{Y}_{i}$ with $\mathcal{X}_{i}=\prod_{k}\mathcal{X}_{i,k}$ such that each subspace contain real matrices : $\mathcal{X}_{i,k}=\mathbb{R}^{n_{i,k}^{X}p_{i,k}^{X}}$ (the same holding for $\mathcal{Y}_{i}$). We also define an ontological index function $I_{X}(i,k)$ (resp. $I_{Y}(i,k)$) taking integer values which coincide if and only if the two variables have the same ontology in the sense of~\cite{livet2010}, i.e. they are supposed to represent the same real object. We distinguish ``raw data'' $\mathbf{X}_{i}$ from which indicators are computed via explicit deterministic functions, from ``intermediate indicators'' $\mathbf{Y}_{i}$ that are already integrated and can be e.g. outputs of elaborated models simulating some aspects of the urban system. We define the partial characteristic space of the ``urban fact'' by 

\begin{equation}
(\mathcal{X},\mathcal{Y}) \underset{def}{=} \left(\prod\tilde{\mathcal{X}}_{c}\right)\times\left(\prod\tilde{\mathcal{Y}}_{c}\right) = \left(\prod_{\mathcal{X}_{i,k}\in\mathcal{D}_{\mathcal{X}}}\mathbb{R}^{p_{i,k}^{X}}\right)\times\left(\prod_{\mathcal{Y}_{i,k}\in\mathcal{D}_{\mathcal{Y}}}\mathbb{R}^{p_{i,k}^{Y}}\right)
\end{equation}

with $\mathcal{D}_{\mathcal{X}}=\{\mathcal{X}_{i,k}|I(i,k)\textrm{ distincts},n_{i,k}^{X}\mbox{ maximal}\}$
(the same holding for $\mathcal{Y}_{i}$). It is indeed the abstract space on which indicators are integrated. The indices $c$ introduced as a definition here correspond to different indicators across all systems. This space is the minimal space common to all systems allowing a common definition for indicators on each.

Let $\mathbf{X}_{i,c}$ be the data canonically projected in the corresponding subspace, well defined for all $i$ and all $c$. We make the key assumption that all indicators are computed by integration against a certain kernel, i.e. that for all $c$, there exists $H_{c}$ space of real-valued functions on $(\tilde{\mathcal{X}}_{c},\tilde{\mathcal{Y}}_{c})$, such that for all $h\in H_{c}$ :
\begin{enumerate}
\item $h$ is ``enough'' regular (tempered distributions e.g.)
\item $q_c=\int_{(\tilde{\mathcal{X}}_{c},\tilde{\mathcal{Y}}_{c})}h$ is a function describing the ``urban fact'' (the indicator in itself)
\end{enumerate}

Typical concrete example of kernels can be :

\begin{itemize}
\item A mean of rows of $\mathbf{X}_{i,c}$ is computed with $h(x)=x\cdot f_{i,c}(x)$ where $f_{i,c}$ is the density of the distribution of the assumed underlying variable.
\item A rate of elements respecting a given condition $C$, $h(x)=f_{i,c}(x)\chi_{C(x)}$ 
\item For already aggregated variables $\mathbf{Y}$, a Dirac distribution allows to express them also as a kernel integral. 
\end{itemize}

\paragraph{Aggregation}

Weighting objectives in multi-attribute decision-making is indeed the crucial point of the processes, and numerous methods are available (see~\cite{wang2009review} for a review for the particular case of sustainable energy management). Let define weights for the linear aggregation. We assume the indicators normalized, i.e. $q_c \in [0,1]$, for a more simple construction of relative weights. For $i,c$ and $h_{c}\in H_{c}$ given, the weight $w_{i,c}$ is simply constituted by the relative importance of the indicator $w_{i,c}^{L}=\frac{\hat{q}_{i,c}}{\sum_{c}\hat{q}_{i,c}}$ where $\hat{q}_{i,c}$ is an estimator of $q_{c}$ for data $\mathbf{X}_{i,c}$ (i.e. the effectively calculated value). Note that this step can be extended to any sets of weight attributions, by taking for example $\tilde{w}_{i,c} = w_{i,c} \cdot w'_{i,c}$ if $\mathbf{w}'$ are the weights attributed by the decision-maker. We focus here on the relative influence of attributes and thus choose this simple form for weights.


\paragraph{Robustness Estimation}

The scene is now set up to be able to estimate the robustness of the evaluation done through the aggregated function. Therefore, we apply an integral approximation method similar to methods introduced in~\cite{varet2010developpement}, since the integrated form of indicators indeed brings the benefits of such powerful theoretical results. Let $\mathbf{X}_{i,c}=(\vec{X}_{i,c,l})_{1\leq l\leq n_{i,c}}$ and $D_{i,c}=Disc_{\tilde{\mathcal{X}}_{c},L^2}(\mathbf{X}_{i,c})$ the discrepancy of data points cloud\footnote{The discrepancy is defined as the $L2$-norm of local discrepancy which is for normalized data points $\mathbf{X}=(x_{ij})\in \left[0,1\right]^d$, a function of $\mathbf{t}\in \left[0,1\right]^d$ comparing the number of points falling in the corresponding hypercube with its volume, by $disc(\mathbf{t}) = \frac{1}{n}\sum_i \mathbbm{1}_{\prod_j x_{ij}<t_j} - \prod_j t_j$. It is a measure of how the point cloud covers the space.}~\cite{niederreiter1972discrepancy}. With $h\in H_{c}$, we have the upper bound on the integral approximation error

\[
\left\Vert \int h_{c}-\frac{1}{n_{i,c}}\sum_{l}h_{c}(\vec{X}_{i,c,l})\right\Vert \leq K\cdot\left|\left|\left|h_{c}\right|\right|\right|\cdot D_{i,c}
\]

where $K$ is a constant independent of data points and objective function. It directly yields

\[
\left\Vert \int\sum w_{i,c}h_{c}-\frac{1}{n_{i,c}}\sum_{l}w_{i,c}h_{c}(\vec{X}_{i,c,l})\right\Vert \leq K\sum_{c}\left|w_{i,c}\right|\left|\left|\left|h_{c}\right|\right|\right|\cdot D_{i,c}
\]

Assuming the error reasonably realized (``worst case'' scenario for knowledge of the theoretical value of aggregated function), we take this upper bound as an approximation of its magnitude. Furthermore, taking normalized indicators implies $\left|\left|\left|h_c\right|\right|\right| = 1$. We propose then to compare error bounds between two evaluations. They depend only on data distribution (equivalent to \emph{statistical robustness}) and on indicators chosen (sort of \emph{ontological robustness}, i.e. do the indicators have a real sense in the chosen context and do their values make sense), and are a way to combine these two type of robustnesses into a single value.

We thus define a \emph{robustness ratio} to compare the robustness of two evaluations by

\begin{equation}
R_{i,i'}=\frac{\sum_{c}w_{i,c}\cdot D_{i,c}}{\sum_{c}w_{i',c}\cdot D_{i',c}}
\end{equation}

The intuitive sense of this definition is that one compares robustness of evaluations by comparing the highest error done in each based on data structure and relative importance.

By taking then an order relation on evaluations by comparing the position of the ratio to one, it is obvious that we obtain a complete order on all possible evaluations. This ratio should theoretically allow to compare any evaluation of an urban system. To keep an ontological sense to it, it should be used to compare disjoints sub-systems with a reasonable proportion of indicators in common, or the same sub-system with varying indicators. Note that it provides a way to test the influence of indicators on an evaluation by analyzing the sensitivity if the ratio to their removal. On the contrary, finding a ``minimal'' number of indicators each making the ratio strongly vary should be a way to isolate essential parameters ruling the sub-system.

\section{Results}


\paragraph{Implementation}

Preprocessing of geographical data is made through QGIS~\cite{qgis2011quantum} for performance reasons. Core implementation of the framework is done in R~\cite{team2000r} for the flexibility of data management and statistical computations. Furthermore, the package \texttt{DiceDesign}~\cite{franco20092} written for numerical experiments and sampling purposes, allows an efficient and direct computation of discrepancies. Last but not least, all source code is openly available on the \texttt{git} repository of the project\footnote{at \url{https://github.com/JusteRaimbault/RobustnessDiscrepancy}} for reproducibility purposes~\cite{ram2013git}.

\subsection{Implementation on Synthetic Data}

We propose in a first time to illustrate the implementation with an application to synthetic data and indicators, for intra-urban quality indicators in the city of Paris.

\paragraph{Data Collection}

We base our virtual case on real geographical data, in particular for \emph{arrondissements} of Paris. We use open data available through the OpenStreetMap project~\cite{bennett2010openstreetmap} that provides accurate high definition data for many urban features. We use the street network and position of buildings within the city of Paris. 
Limits of \emph{arrondissements}, used to overlay and extract features when working on single districts, are also extracted from the same source. We use centroids of buildings polygons, and segments of street network. Dataset overall consists of around $200k$ building features and $100k$ road segments.

\paragraph{Virtual Cases}

We work on each district of Paris (from the 1st to the 20th) as an evaluated urban system. We construct random synthetic data associated to spatial features, so each district has to be evaluated many time to obtain mean statistical behavior of toy indicators and robustness ratios. The indicators chosen need to be computed on residential and street network spatial data. We implement two mean kernels and a conditional mean to show different examples, linked to environmental sustainability and quality of life, that are required to be maximized. Note that these indicators have a real meaning but no particular reason to be aggregated, they are chosen here for the convenience of the toy model and the generation of synthetic data. With $a\in \{1\ldots 20\}$ the number of the district, $A(a)$ corresponding spatial extent, $b\in B$ building coordinates and $s\in S$ street segments, we take
\begin{itemize}
\item Complementary of the average daily distance to work with car per individual, approximated by, with $n_{cars}(b)$ number of cars in the building (randomly generated by associated of cars to a number of building proportional to motorization rate $\alpha_m ~ 0.4$ in Paris), $d_w$ distance to work of individuals (generated from the building to a uniformly generated random point in spatial extent of the dataset), and $d_{max}$ the diameter of Paris area, $\bar{d}_w = 1 - \frac{1}{|b\in A(a)|} \cdot \sum_{b\in A(a)}{n_{cars}(b)\cdot \frac{d_w}{d_{max}}}$
\item Complementary of average car flows within the streets in the district, approximated by, with $\varphi(s)$ relative flow in street segment $s$, generated through the minimum of 1 and a log-normal distribution adjusted to have $95\%$ of mass smaller than 1 what mimics the hierarchical distribution of street use (corresponding to betweenness centrality), and $l(s)$ segment length, $\bar{\varphi} = 1 - \frac{1}{|s\in A(a)|} \cdot \sum_{s \in A(a)}{\varphi(s)\cdot \frac{l(s)}{\max{(l(s))}}}$

\item Relative length of pedestrian streets $\bar{p}$, computed through a randomly uniformly generated dummy variable adjusted to have a fixed global proportion of segments that are pedestrian.
\end{itemize}


\begin{table}[h!]
\hspace{-1cm}
\begin{tabular}[6pt]{c|c|c|c|c}
Arrdt & $<\bar{d}_w> \pm \sigma (\bar{d}_w)$ & $<\bar{\varphi}> \pm \sigma (\bar{\varphi})$ & $<\bar{p}> \pm \sigma (\bar{p})$ & $R_{i,1}$ \\[3pt]
\hline
1 th & 0.731655 $\pm$ 0.041099 & 0.917462 $\pm$ 0.026637 & 0.191615 $\pm$ 0.052142 & 1.000000 $\pm$ 0.000000\\[3pt]
\hline
2 th & 0.723225 $\pm$ 0.032539 & 0.844350 $\pm$ 0.036085 & 0.209467 $\pm$ 0.058675 & 1.002098 $\pm$ 0.039972\\[3pt]
\hline
3 th & 0.713716 $\pm$ 0.044789 & 0.797313 $\pm$ 0.057480 & 0.185541 $\pm$ 0.065089 & 0.999341 $\pm$ 0.048825\\[3pt]
\hline
4 th & 0.712394 $\pm$ 0.042897 & 0.861635 $\pm$ 0.030859 & 0.201236 $\pm$ 0.044395 & 0.973045 $\pm$ 0.036993\\[3pt]
\hline
5 th & 0.715557 $\pm$ 0.026328 & 0.894675 $\pm$ 0.020730 & 0.209965 $\pm$ 0.050093 & 0.963466 $\pm$ 0.040722\\[3pt]
\hline
6 th & 0.733249 $\pm$ 0.026890 & 0.875613 $\pm$ 0.029169 & 0.206690 $\pm$ 0.054850 & 0.990676 $\pm$ 0.031666\\[3pt]
\hline
7 th & 0.719775 $\pm$ 0.029072 & 0.891861 $\pm$ 0.026695 & 0.209265 $\pm$ 0.041337 & 0.966103 $\pm$ 0.037132\\[3pt]
\hline
8 th & 0.713602 $\pm$ 0.034423 & 0.931776 $\pm$ 0.015356 & 0.208923 $\pm$ 0.036814 & 0.973975 $\pm$ 0.033809\\[3pt]
\hline
9 th & 0.712441 $\pm$ 0.027587 & 0.910817 $\pm$ 0.015915 & 0.202283 $\pm$ 0.049044 & 0.971889 $\pm$ 0.035381\\[3pt]
\hline
10 th & 0.713072 $\pm$ 0.028918 & 0.881710 $\pm$ 0.021668 & 0.210118 $\pm$ 0.040435 & 0.991036 $\pm$ 0.038942\\[3pt]
\hline
11 th & 0.682905 $\pm$ 0.034225 & 0.875217 $\pm$ 0.019678 & 0.203195 $\pm$ 0.047049 & 0.949828 $\pm$ 0.035122\\[3pt]
\hline
12 th & 0.646328 $\pm$ 0.039668 & 0.920086 $\pm$ 0.019238 & 0.198986 $\pm$ 0.023012 & 0.960192 $\pm$ 0.034854\\[3pt]
\hline
13 th & 0.697512 $\pm$ 0.025461 & 0.890253 $\pm$ 0.022778 & 0.201406 $\pm$ 0.030348 & 0.960534 $\pm$ 0.033730\\[3pt]
\hline
14 th & 0.703224 $\pm$ 0.019900 & 0.902898 $\pm$ 0.019830 & 0.205575 $\pm$ 0.038635 & 0.932755 $\pm$ 0.033616\\[3pt]
\hline
15 th & 0.692050 $\pm$ 0.027536 & 0.891654 $\pm$ 0.018239 & 0.200860 $\pm$ 0.024085 & 0.929006 $\pm$ 0.031675\\[3pt]
\hline
16 th & 0.654609 $\pm$ 0.028141 & 0.928181 $\pm$ 0.013477 & 0.202355 $\pm$ 0.017180 & 0.963143 $\pm$ 0.033232\\[3pt]
\hline
17 th & 0.683020 $\pm$ 0.025644 & 0.890392 $\pm$ 0.023586 & 0.198464 $\pm$ 0.033714 & 0.941025 $\pm$ 0.034951\\[3pt]
\hline
18 th & 0.699170 $\pm$ 0.025487 & 0.911382 $\pm$ 0.027290 & 0.188802 $\pm$ 0.036537 & 0.950874 $\pm$ 0.028669\\[3pt]
\hline
19 th & 0.655108 $\pm$ 0.031857 & 0.884214 $\pm$ 0.027816 & 0.209234 $\pm$ 0.032466 & 0.962966 $\pm$ 0.034187\\[3pt]
\hline
20 th & 0.637446 $\pm$ 0.032562 & 0.873755 $\pm$ 0.036792 & 0.196807 $\pm$ 0.026001 & 0.952410 $\pm$ 0.038702\\[3pt]
\hline
\end{tabular}

\bigskip

\caption{Numerical results of simulation for each district with $N=50$ repetitions. Each toy indicator value is given by mean on repetitions and associated standard deviation. Robustness ratio is computed relative to first district (arbitrary choice). A ratio smaller than 1 means that integral bound is smaller for upper district, i.e. that evaluation is more robust for this district. Because of the small size of first district, we expected a majority of district to give ratio smaller than 1, what is confirmed by results, even when adding standard deviations.}

\end{table}

As synthetic data are stochastic, we run the computation for each district $N=50$ times, what was a reasonable compromise between statistical convergence and time required for computation. Table 1 shows results (mean and standard deviations) of indicator values and robustness ratio computation. Obtained standard deviation confirm that this number of repetitions give consistent results. Indicators obtained through a fixed ratio show small variability what may a limit of this toy approach. However, we obtain the interesting result that a majority of districts give more robust evaluations than 1st district, what was expected because of the size and content of this district : it is indeed a small one with large administrative buildings, what means less spatial elements and thus a less robust evaluation following our definition of the robustness.

\subsection{Application to a Real Case : Metropolitan Segregation}

The first example was aimed to show potentialities of the method but was purely synthetic, hence yielding no concrete conclusion nor implications for policy. We propose now to apply it to real data for the example of metropolitan segregation.

\paragraph{Data}

We work on income data available for France at an intra-urban level (basic statistical units IRIS) for the year 2011 under the form of summary statistics (deciles if the area is populated enough to ensure anonymity), provided by INSEE\footnote{\texttt{http://www.insee.fr}}. Data are associated with geographical extent of statistical units, allowing computation of spatial analysis indicators. 

\paragraph{Indicators}

We use here three indicators of segregation integrated on a geographical area. Let assume the area divided into covering units $\mathcal{S}_i$ for $1\leq i \leq N$ with centroids $(x_i,y_i)$. Each unit has characteristics of population $P_i$ and median income $X_i$. We define spatial weights used to quantify strength of geographical interactions between units $i,j$, with $d_{ij}$ euclidian distance between centroids : $w_{ij} = \frac{P_i P_j}{\left(\sum_k P_k\right)^2}\cdot \frac{1}{d_{ij}}$ if $i\neq i$ and $w_{ii} = 0$. The normalized indicators are the following

\begin{itemize}
\item Spatial autocorrelation Moran index, defined as weighted normalized covariance of median income by $\rho = \frac{N}{\sum_{ij}w_{ij}}\cdot \frac{\sum_{ij}w_{ij}\left(X_i - \bar{X}\right)\left(X_j - \bar{X}\right)}{\sum_i \left(X_i - \bar{X}\right)^2}$
\item Dissimilarity index (close to Moran but integrating local dissimilarities rather than correlations), given by $d =  \frac{1}{\sum_{ij}w_{ij}} \sum_{ij} w_{ij} \left|\tilde{X}_i - \tilde{X}_j\right|$\\ with $\tilde{X}_i = \frac{X_i - \min(X_k)}{\max(X_k) - \min(X_k)}$
\item Complementary of the entropy of income distribution that is a way to capture global inequalities $\varepsilon = 1 + \frac{1}{\log(N)} \sum_i \frac{X_i}{\sum_k X_k} \cdot \log\left(\frac{X_i}{\sum_k X_k}\right)$
\end{itemize}

Numerous measures of segregation with various meanings and at different scales are available, as for example at the level of the unit by comparison of empirical wage distribution with a theoretical null model~\cite{louf2015patterns}. The choice here is arbitrary in order to illustrate our method with a reasonable number of dimensions.

\begin{figure}[h!]
\centering
\hspace{-3cm}\includegraphics[width=1.2\textwidth]{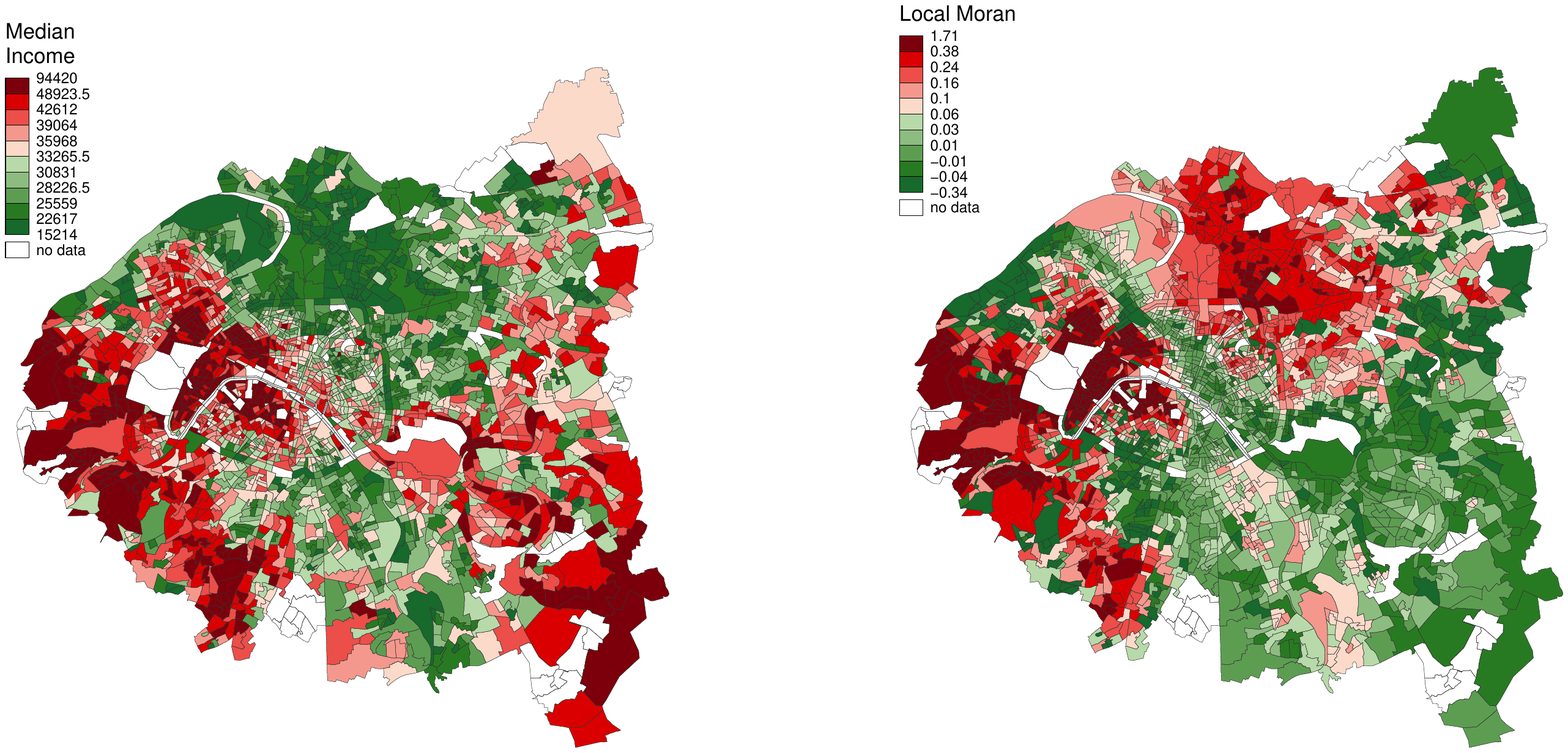}\hspace{-2cm}
\vspace{-2.5cm}
\caption{\textbf{Maps of Metropolitan Segregation.} Maps show yearly median income on basic statistical units (IRIS) for the three departments constituting mainly the Great Paris metropolitan area, and the corresponding local Moran spatial autocorrelation index, defined for unit $i$ as $\rho_i = N/\sum_{j}w_{ij} \cdot \frac{\sum_{j} w_{ij} (X_j - \bar{X})(X_i - \bar{X})}{\sum_i (X_i - \bar{X})^2}$. The most segregated areas coincide with the richest and the poorest, suggesting an increase of segregation in extreme situations.}
\end{figure}

\paragraph{Results}

We apply our method with these indicators on the Greater Paris area, constituted of four \emph{d{\'e}partements} that are intermediate administrative units. The recent creation of a new metropolitan governance system~\cite{gilli2009paris} underlines interrogations on its consistence, and in particular on its relation to intermediate spatial inequalities. We show in Fig. 1 maps of spatial distribution of median income and corresponding local index of autocorrelation. We observe the well-known West-East opposition and district disparities inside Paris as they were formulated in various studies, such as~\cite{guerois2009dynamique} through the analysis of real estate transactions dynamics. We then apply our framework to answer a concrete question that has implications for urban policy : \textit{how are the evaluation of segregation within different territories sensitive to missing data ?} To do so, we proceed to Monte Carlo simulations (75 repetitions) during which a fixed proportion of data is randomly removed, and the corresponding robustness index is evaluated with renormalized indicators. Simulations are done on each \emph{department} separately, each time relatively to the robustness of the evaluation of full Greater Paris. Results are shown in Fig. 2. All areas present a slightly better robustness than the reference, what could be explained by local homogeneity and thus more fiable segregation values. Implications for policy that can be drawn are for example direct comparisons between areas : a loss of 30\% of information on 93 area corresponds to a loss of only 25\% in 92 area. The first being a deprived area, the inequality is increased by this relative lower quality of statistical information. The study of standard deviations suggest further investigations as different response regimes to data removal seem to exist.

\begin{figure}
\centering
\hspace{-2cm}\includegraphics[width=0.55\textwidth]{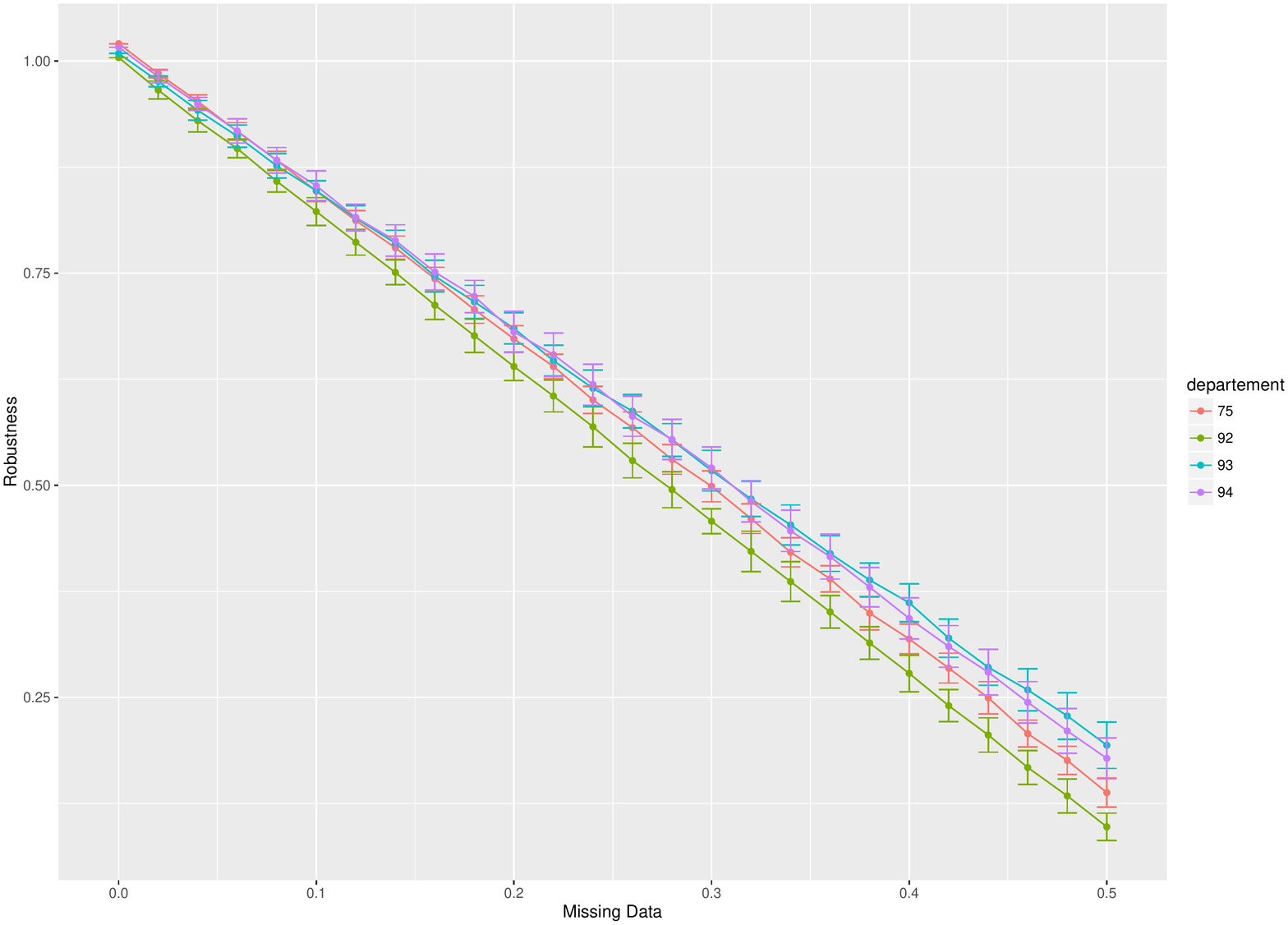}
\includegraphics[width=0.55\textwidth]{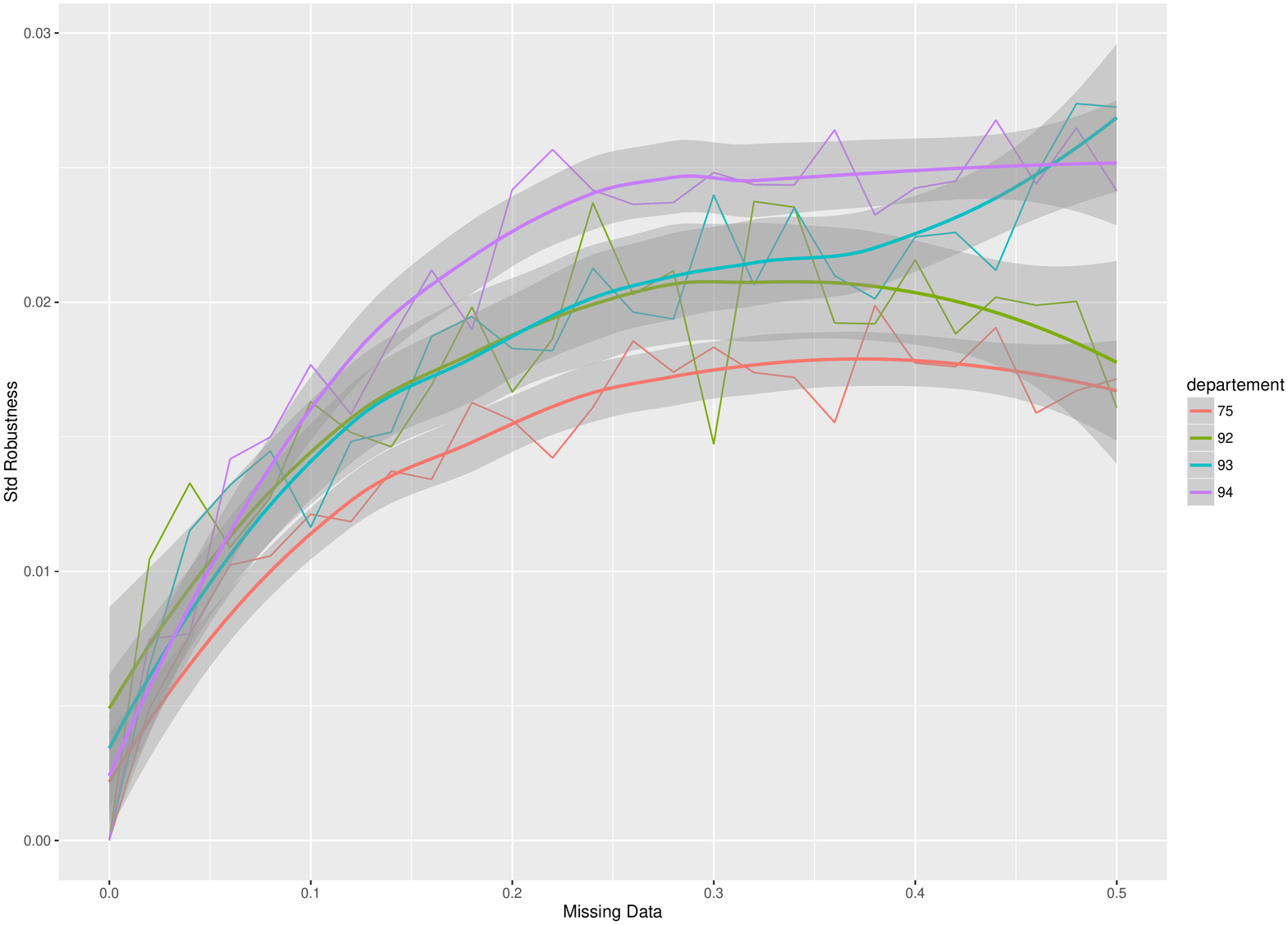}\hspace{-2cm}
\caption{\textbf{Sensitivity of robustness to missing data.} \textit{Left.} For each department, Monte Carlo simulations (N=75 repetitions) are used to determine the impact of missing data on robustness of segregation evaluation. Robustness ratios are all computed relatively to full metropolitan area with all available data. Quasi-linear behavior translates an approximative linear decrease of discrepancy as a function of data size. The similar trajectory of poorest departments (93,94) suggest the correction to linear behavior being driven be segregation patterns. \textit{Right.} Corresponding standard deviations of robustness ratios. Different regimes (in particular 93 against others) unveil phase transitions at different levels of missing data, meaning that the evaluation in 94 is from this point of view more sensitive to missing data.}
\end{figure}

\section{Discussion}

\subsection{Applicability to Real situations}

\paragraph{Implications for Decision-making}

The application of our method to concrete decision-making can be thought in different ways. First in the case of a comparative multi-attribute decision process, such as the determination of a transportation corridor, the identification of territories on which the evaluation may be flawed (i.e. has a poor relative robustness) could allow a more refined focus on these and a corresponding revision of datasets or an adapted revision of weights. In any case the overall decision-making process should be made more reliable. A second direction lays in the spirit of the real application we have proposed, i.e. the sensitivity of evaluation to various parameters such as missing data. If a decision appears as reliable because data have few missing points, but the evaluation is very sensitive to it, one will be more careful in the interpretation of results and taking the final decision. Further work and testing will however be needed to understand framework behavior in different contexts and be able to pilot its application in various real situations.

\paragraph{Integration Within Existing Frameworks}

The applicability of the method on real cases will directly depend on its potential integration within existing framework. Beyond technical difficulties that will surely appear when trying to couple or integrate implementations, more theoretical obstacles could occur, such as fuzzy formulations of functions or data types, consistency issues in databases, etc. Such multi-criteria framework are numerous. Further interesting work would be to attempt integration into an open one, such as e.g. the one described in~\cite{tivadar2014oasis} which calculates various indices of urban segregation, as we have already illustrated the application on metropolitan segregation indexes.

\paragraph{Availability of Raw Data}

In general, sensitive data such as transportation questionnaires, or very fine granularity census data are not openly available but provided already aggregated at a certain level (for instance French Insee Data are publicly available at basic statistical unit level or larger areas depending on variables and minimal population constraints, more precise data is under restricted access). It means that applying the framework may imply complicated data research procedure, its advantage to be flexible being thus reduced through additional constraints.

\subsection{Validity of Theoretical Assumptions}

A possible limitation of our approach is the validity of the assumption formulating indicators as spatial integrals. Indeed, many socio-economic indicators are not necessarily depending explicitly on space, and trying to associate them with spatial coordinates may become a slippery slope (e.g. associate individual economic variables with individual residential coordinates will have a sense only if the use of the variable has a relation with space, otherwise it is a non-legitimate artifact). Even indicators which have a spatial value may derive from non-spatial variables, as~\cite{kwan1998space} points out concerning accessibility, when opposing integrated accessibility measures with individual-based non necessarily spatial-based (e.g. individual decisions) measures. Constraining a theoretical representation of a system to fit a framework by changing some of its ontological properties (always in the sense of real meaning of objects) can be understood as a violation of a fundamental rule of modeling and simulation in social science given in~\cite{banos2013HDR}, that is that there can be an universal ``language'' for modeling and some can not express some systems, having for consequence misleading conclusion due to ontology breaking in the case of an over-constrained formulation.

\subsection{Framework Generality}

We argue that the fundamental advantage of the proposed framework is its generality and flexibility, since robustness of the evaluations are obtained only through data structure if ones relaxes constraints on the value of weight. Further work should go towards a more general formulation, suppressing for example the linear aggregation assumption. Non-linear aggregation functions would require however to present particular properties regarding integral inequalities. For example, similar results could search in the direction of integral inequalities for Lipschitzian functions such as the one-dimensional results of~\cite{dragomir1999ostrowski}.

\section*{Conclusion}

We have proposed a model-independent framework to compare the robustness of multi-attribute evaluations between different urban systems. Based on data discrepancy, it provide a general definition of relative robustness without any assumption on model for the system, but with limiting assumptions that are the need of linear aggregation and of indicators being expressed through spatial kernel integrals. We propose a toy implementation based on real data for the city of Paris, numerical results confirming general expected behavior, and an implementation on real data for income segregation on Greater Paris metropolitan areas, giving possible insights into concrete policy questions. Further work should be oriented towards sensitivity analysis of the method, application to other real cases and theoretical assumptions relaxation, i.e. the relaxation of linear aggregation and spatial integration.

\section*{Acknowledgments}

The author would like to thank Julien Keutchayan (Ecole Polytechnique de Montr{\'e}al) for suggesting the original idea of using discrepancy, and anonymous reviewers for the useful comments and insights.


\bibliographystyle{unsrt}
\bibliography{biblio.bib}

\end{document}